\documentclass[reqno]{amsart}
\usepackage{amssymb,graphics,graphicx,bbm, float}

\def\be{\begin{equation}}
\def\ee{\end{equation}}
\def\bm{\begin{multline}}
\def\bfig{\begin{figure}[htb]}
\def\efig{\end{figure}}

\numberwithin{equation}{section}
\newtheorem{theorem}{Theorem}[section]
\newtheorem{proposition}[theorem]{Proposition}

\newtheorem{corollary}[theorem]{Corollary}
\newtheorem{assumption}{Assumption}
\newtheorem{definition}{Definition}
\newtheorem{remark}{Remark}

\begin{document}

\title{Virial Expansion Bounds}

\author[S. J. Tate]{Stephen James Tate}
\address{Department of Mathematics, University of Warwick,
Coventry, CV4 7AL, United Kingdom}
\email{s.j.tate@warwick.ac.uk}



\begin{abstract}
In the 1960s, the technique of using cluster expansion bounds in order to achieve bounds on the virial expansion was developed by Lebowitz and Penrose (1964) and Ruelle (1969). This technique is generalised to more recent cluster expansion bounds by Poghosyan and Ueltschi (2009), which are related to the work of Procacci (2007) and the tree-graph identity, detailed by Brydges (1986). The bounds achieved by Lebowitz and Penrose can also be sharpened by doing the actual optimisation and achieving expressions in terms of the Lambert W-function. The different bound from the cluster expansion shows some improvements for bounds on the convergence of the virial expansion in the case of positive potentials, which are allowed to have a hard core. 
\end{abstract}

\thanks{Work partially supported by EPSRC grant EP/G056390/1}
\thanks{\copyright{} 2013 by the author. This paper may be reproduced, in its
entirety, for non-commercial purposes.}

\maketitle

\section{Introduction}
\label{sec: intro}
In this article, further bounds on the virial coefficients and consequently the radius of convergence of the virial expansion are developed in a similar way to how they are achieved in the paper by Lebowitz and Penrose \cite{LPen64}, that is via cluster expansion bounds. These techniques are shown in general and they rely upon the Lambert W-function. They give bounds which improve upon those given by Lebowitz and Penrose and they are applied to a more recent version of cluster expansion bounds, detailed by Poghosyan and Ueltschi \cite{PU09}. The new bounds display an improvement for small temperatures (large $\beta$) and potentials with a hard core and sufficiently attractive tail. There is no difference for purely repulsive potentials.
\subsection{Background to the Paper}
The bounds on the virial coefficients and consequently, the radius of convergence of the virial expansion, were mainly achieved in the 1960s by Lebowitz and Penrose \cite{LPen64}, Groeneveld \cite{G62} and Ruelle \cite{R69}. These bounds have not been improved upon in the literature, even though our understanding of cluster and virial expansions has improved since then. Moreover, further, more refined, techniques have been developed, which mainly focus on tree-graph identities. Recent work by Pulvirenti and Tsagkarogiannis \cite{PT12} and Morais and Procacci \cite{MP13} have made some gains in understanding bounds on coeffcients, through the use of Canonical Ensemble calculations, which naturally use the parameter of density, rather than the fugacity, which is the parameter in the Grand Canonical Ensemble. This saves needing to invert the fugacity as a function of density or using Lagrange inversion ideas. Furthermore, the recent paper by Jansen \cite{J12} raised important questions about achieving the low temperature asymptotics, which are of the form: $e^{-\beta B}$, rather than $e^{-2\beta B}$, which encourages a search for an improvement to the current bounds.
\section{Main Results}
The main results of this paper rely on the particle model of the classical gas. 
The main parameters in the model are: the inverse temperature $\beta = \frac{1}{kT}$, where $k$ is Boltzmann's constant and $T$ is the temperature; the fugacity $z$; the pressure $P$; and the density $\rho$.

We start with the Grand Canonical Partition function for a classical gas:
\be \Xi(z)  = \sum\limits_{N=0}^{\infty}Z_Nz^n \ee
Where $Z_N$ is the $N$-particle configuration integral:
\be Z_N = \frac{1}{N!}\prod\limits_{i=1}^N\left(\int_{\mathbb{R}^d}\, \mathrm{d}^dx_i\right) e^{-\beta\sum\limits_{1 \leq i<j \leq N}\Phi(x_i,x_j)} \ee
We have the relationship $\beta P = \ln \Xi$, which gives us an expansion for pressure. 
The power series for pressure in terms of fugacity $z$, may be written as:
\be \beta P = \sum\limits_{n \geq 1}b_nz^n \label{clusterexp} \ee
Using the formula $\rho=z\frac{\partial}{\partial z} (\beta P)$, we have the corresponding fugacity expansion for density:
\be \rho = \sum\limits_{n \geq 1} nb_n z^n \label{clusterexp2} \ee
We see that the power series expansion for $\rho$ has a zero constant term and non zero $z$ term, where we have that $b_1 \neq 0$ and usually we set it to $1$. It is usual to have a non zero $z$ coefficient in the cluster expansion. This means that it is possible to invert the $\rho-z$ relationship, in order to obtain an expansion for $z(\rho)$. This can be substituted into \eqref{clusterexp} in order to get a power series for $P$ in terms of $\rho$:
\be P=\sum\limits_{n \geq 1}c_n \rho^n \label{virialexp} \ee
The assumptions for the initial cluster expansion bounds used in this article are:
\begin{assumption}[Potential]
The $N$-particle interaction potential: $U_N(x_1, \cdots, x_N)$ may be written as the sum of pair-potentials:
\be U_N(x_1, \cdots, x_N)= \sum\limits_{1 \leq i<j \leq N}\Phi(x_i,x_j) \ee
Furthermore, we assume that the pair potentials $\Phi(x_i,x_j)$ are central, that is, they only depend on the distance from $x_i$ to $x_j$.
\end{assumption}
\begin{assumption}[Stability]
\label{ass: stability}
The potential energy is assumed to be stable, that is, there is a $B>0$, such that for every $N$ and $(x_1, \cdots, x_N) \in \mathbb{R}^{Nd}$, we have:
\be U(x_1, \cdots, x_N)=\sum\limits_{1 \leq i<j \leq N}\Phi(x_i,x_j) \geq -BN \ee
where $d$ is the dimension of our system.
\end{assumption}
\begin{definition}[$C(\beta)$ and $R(\beta)$ - `temperedness']
We have two main functions of $\beta$, which play an important r\^{o}le in the cluster expansion bounds:
\be C(\beta):=\int\limits_{\mathbb{R}^d} \, \left|e^{-\beta \Phi(0,x)}-1\right| \mathrm{d}^dx \label{Cbeta} \ee
\be R(\beta):= \left(|\mathbb{B}|r^d+\beta \int\limits_{|y|>r}\, |\Phi(0,y)| \mathrm{d}^dy \right) \label{Rbeta} \ee
If the expression \eqref{Cbeta} is finite then the potential $\Phi$ is called `tempered', which we assume for bounds involving $C(\beta)$

The $r$ in \eqref{Rbeta} represents the radius of the hard-core interaction. $|\mathbb{B}|$ is the surface area of a $d$-dimensional sphere. 
\end{definition}
We write $\mathcal{R}_{\text{Vir}}$ for the radius of convergence of the virial expansion.
\begin{definition}[Lambert W-function]
We denote by $W(z)$, the Lambert W-function with domain $\mathbb{R}_+:=\{x \in \mathbb{R} \vert x \geq 0\}$ and range $\mathbb{R}_+$. It is the solution to:
\be W(z) e^{W(z)}=z \ee
\end{definition}

Further notes on the Lambert W-function can be found in \cite{CJK97}

\begin{theorem}[General Virial Bounds]
\label{thm: general}
Assuming cluster coefficient bounds of the form:
\be |nb_n| \leq a \frac{n^{n-1}}{n!}b^n \ee
where $a$ and $b$ are non-negative functions of inverse temperature $\beta$, we have the virial coefficient bounds:
\be |c_n| \leq \frac{\beta^{-1}}{n}\left(a^{-1} \frac{W\left(\frac{eab}{1+ab}\right)}{\left(W\left(\frac{eab}{1+ab}\right)-1\right)^2}\right)^{n-1} \ee
which gives the lower bound on the radius of convergence as:
\be \mathcal{R}_{\text{Vir}} \geq a \frac{\left(W\left(\frac{eab}{1+ab}\right)-1\right)^2}{W\left(\frac{eab}{1+ab}\right)} \ee
\end{theorem} 
If we apply this general theorem, which is derived in Section \ref{sec: general}, to two specific bounds we have for cluster expansions, we achieve:
\begin{corollary}[Improved Lebowitz-Penrose]
\label{cor:lebpenbound}
The cluster expansion bounds:
\be |nb_n| \leq \frac{n^{n-1}}{n!}e^{2\beta B(n-1)}C(\beta)^{n-1} \ee
give the bound for virial coefficients:
\be |c_n| \leq \frac{\beta^{-1}}{n} \left(C(\beta)e^{4\beta B} \frac{W\left(\frac{e}{1+e^{2\beta B}}\right)}{\left(W\left(\frac{e}{1+e^{2\beta B}}\right)-1\right)^2}\right)^{n-1} \ee
and the bound for the radius of convergence:
\be \mathcal{R}_{\text{Vir}} \geq C(\beta)^{-1} e^{-4\beta B}\frac{\left(W\left(\frac{e}{1+e^{2\beta B}}\right)-1\right)^2}{W\left(\frac{e}{1+e^{2\beta B}}\right)} \ee
\end{corollary}
This is shown in Section \ref{sec: oldbounds}

For purely hard-core interactions $B=0$ and the radius of convergence satisfies:
\be \mathcal{R}_{\text{Vir}} \geq C(\beta)^{-1} \frac{\left(W\left(\frac{e}{2}\right)-1\right)^2}{W\left(\frac{e}{2}\right)} \ee
This is precisely the same as what is obtained by Lebowitz-Penrose.

\begin{corollary}[Alternative Bounds]
\label{cor:mybounds}
For cluster expansion bounds:
\be |nb_n| \leq \frac{n^{n-1}}{n!}R(\beta)^{n-1}e^{n\beta B} \ee
we have the bound for virial coefficients as:
\be |c_n| \leq \frac{\beta^{-1}}{n}R(\beta)^{n-1} \left( \frac{W\left(\frac{e^{\beta B+1}}{1+e^{\beta B}}\right)}{\left(W\left(\frac{e^{\beta B +1}}{1+e^{\beta B}}\right)-1\right)^2}\right)^{n-1} \ee
Thus giving the radius of convergence as:
\be \mathcal{R}_{\text{Vir}} \geq R(\beta)^{-1} \frac{\left(W\left(\frac{e^{\beta B +1}}{1+e^{\beta B}}\right)-1\right)^2}{W\left(\frac{e^{\beta B+1}}{1+e^{\beta B}}\right)} \ee
\end{corollary}
\begin{remark}
The bounds for the cluster coefficients used in Corollary \ref{cor:lebpenbound} are obtained in \cite{R69} and \cite{LPen64}.  
The bounds for the cluster coefficients used in Corollary \ref{cor:mybounds} are obtained in \cite{PU09}.
\end{remark}
Having a new separate bound is only a good idea, if it is an improvement in certain cases. In Section \ref{sec: comparison} we understand how the two different bounds compare and conclude:
\begin{proposition}[Comparison of Bounds]
\label{prop: comparison}
The bound in Corollary \ref{cor:mybounds} is better than that in Corollary \ref{cor:lebpenbound}, precisely when 
\be 1.6R(\beta) < C(\beta) \ee
\end{proposition}
This is likely to happen at low temperatures (large $\beta$) and for potentials with a hard core and attractive tail. The attractive element of the potential is important, since the improvements are made corresponding to the stability parameter $B$.

\section{Obtaining the Relationship}

Lagrange Inversion techniques are the usual approach to obtaining the virial coefficients and this has a nice representation as a contour integral. If we want to compute the coefficient $c_n$, we can take the contour integral (around $0$) of $\frac{\partial P}{\partial \rho}$ divided by $n\rho^n$. This gives us the formula:
\be c_n = \frac{1}{2\pi i}\oint_C \, \frac{\frac{\partial P}{\partial \rho}}{n\rho^n}\mathrm{d}\rho \ee
We can manipulate this equation to get it in terms of the $z$-variable, since we know about the cluster expansion already.
\begin{align} c_n &= \frac{1}{2 \pi i} \oint_{C'} \, \frac{\frac{\partial P}{\partial z}}{n\rho^n}\mathrm{d}z \notag \\
&= \frac{\beta^{-1}}{2n \pi i} \oint_{C'} \, \frac{z\frac{\partial \beta P}{\partial z}}{z\rho^n} \mathrm{d}z \notag \\
&= \frac{\beta^{-1}}{2n \pi i} \oint_{C'} \, \frac{\mathrm{d}z}{z\rho^{n-1}} \label{contourrep} \end{align}
This (implicit) relationship between the virial and the cluster coefficients is our starting point. The idea is that we can bound such an integral quite easily when we consider bounds on $|\rho|$, when we write it in terms of the fugacity $z$ as in \eqref{clusterexp2}.

Indeed this method is the one used by Ruelle \cite{R69} and Lebowitz and Penrose \cite{LPen64}. 

\section{General Derivation: Derivation for Theorem \ref{thm: general}}
\label{sec: general}
We wish to obtain bounds on virial coefficients using those which come from cluster coefficients, in order to gain an estimate of the lower bound for the radius of convergence of the virial expansion. The derivation loosely follows the method used by Ruelle in \cite{R69} and that of Lebowitz and Penrose in \cite{LPen64}. The starting point is the cluster expansion bound, which is given in the form:
\be |nb_n| \leq a \frac{n^{n-1}}{n!}b^n \label{generalcluster} \ee
Where $a$ and $b$ are positive functions of temperature defined by the particular bound we use.

The starting point is the relationship in \eqref{contourrep}, (implicitly) defining virial coefficients in terms of cluster coefficients:
\be c_n=\frac{\beta^{-1}}{2 \pi i} \oint_C \frac{\mathrm{d}z}{nz\rho^{n-1}} \label{integralforvirial} \ee
In order to get an upper bound on $|c_n|$, we need to bound $|\rho|$ from below. To do this we use the bound:
\be |\rho-z| \leq \sum\limits_{n=2}^{\infty} |nb_n||z|^n \label{r-zbound} \ee
We then substitute in for the upper bounds we have on the cluster coefficients:
\be |\rho-z| \leq a \sum\limits_{n=2}^{\infty}\frac{n^{n-1}}{n!}(b|z|)^n \label{r-zbound2} \ee
We define the function: $f(x):=\sum\limits_{n=1}^{\infty}\frac{n^{n-1}}{n!}x^n$ and write \eqref{r-zbound2} conveniently as:
\be |\rho-z| \leq a(f(b|z|)-b|z|) \label{simplebound} \ee
Using the reverse triangle inequality, we obtain:
\be |\rho| \geq |z|(1+ab)-a f(b|z|) \label{rbound} \ee
Let $b|z|=se^{-s}$ and observe that, from the assumed generic bound on the cluster expansion, $b|z| \leq e^{-1}$ in order for $\sum\limits_{n=1}^{\infty}nb_nz^n$ to converge. $se^{-s}$ is an increasing function on $(0,1)$ and it takes values in $(0,e^{-1})$ as required. We then have the bound in terms of $s$:
\be |\rho|\geq b^{-1}se^{-s}(1+ab)-af(se^{-s}) \label{sfunction1} \ee
We have chosen the function $se^{-s}$ since it is the inverse function of $f$. This can be understood from Lagrange Inversion (shown in appendix) and is a result in \cite{CJK97} . We thus have the expression:
\be |\rho| \geq b^{-1}se^{-s}(1+ab)-as \label{sfunction2} \ee
We thus seek to maximise the right hand side to get the best possible bound. We notice that in the range $s \in (0,1)$ we have a zero at $s=0$ and another when $b^{-1}e^{-s}(1+ab)-a=0$ i.e. when $e^{s}=1+\frac{1}{ab}$, so at $s=\ln\left(1+\frac{1}{ab}\right)  $. It is positive for $s \in (0,\ln\left(1+\frac{1}{ab}\right))$. We seek the value of $s$ to maximise this function in this range.
\begin{remark}
This approach to estimating a lower bound for the radius of convergence for the virial expansion, takes a value of $|z|$ some distance away from its maximal value for convergent cluster expansions. We expect density to increase with fugacity and so the maximal density for which the cluster expansions remain convergent appears to be greater than that for the virial expansion. This is not true in general and is a weakness of this approach.
\end{remark}
If we define: 
\be r(s):=s(e^{-s}(1+ab)b^{-1}-a) \label{reqn} \ee
and take the derivative to search for an extremum in this range.
\be r'(s) = (e^{-s}(1+ab)b^{-1}-a)-se^{-s}(1+ab)b^{-1} \label{r'eqn} \ee
if we find when $r'(s)=0$ in \eqref{r'eqn}, then we solve:
\be (1-s)e^{-s}=\frac{ab}{1+ab} \ee
Substituting $\gamma=1-s$, we get the equation for $\gamma$:
\be \gamma e^{\gamma}=\frac{eab}{1+ab} \label{lambertfunction} \ee
The Lambert $W$-function, as is explained in \cite{CJK97}, is the inverse of $\gamma e^{\gamma}$ and so we can write \eqref{reqn} in terms of $\gamma$ and substitute $\gamma$ for $W(\mu)$, where $\mu := \frac{eab}{1+ab}$. If we write $\tilde{r}(\gamma(s)):=r(s)$. We thus get:
\begin{align} \tilde{r}(\gamma)&=(1-\gamma)(e^{\gamma}\frac{1+ab}{eb}-a) \notag \\
 &= a(e^{\gamma}\frac{1+ab}{eab}-\gamma e^{\gamma} \frac{1+ab}{eab}-1+\gamma) \notag \\
 &=a(\frac{1}{\gamma}-2+\gamma) \notag \\
 &=a\frac{(W(\mu)-1)^2}{W(\mu)} \label{finalreqn}
\end{align}
Where we use \eqref{lambertfunction} to cancel: $\gamma e^{\gamma} \frac{1+ab}{eab}=1$ and $e^{\gamma} \frac{1+ab}{eab}=\frac{1}{\gamma}$.

We evaluate the integral \eqref{integralforvirial} along the contour described by the circle $|z|=$ constant, where the constant is determined by the manipulations above. This leaves us with the integral: 
\be |c_n| \leq \frac{\beta^{-1}}{2n\pi} \oint_C \frac{\mathrm{d}z}{|z||\rho|^{n-1}} \ee
This gives us bounds on the coefficients $c_n$ as:
\be |c_n| \leq \frac{\beta^{-1}}{n} \left(a^{-1}\frac{W(\mu)}{(W(\mu)-1)^2}\right)^{n-1} \label{cbounds} \ee

This gives us the radius of convergence of the virial expansion satisfying:
\be \mathcal{R}_{\text{Vir}} \geq a \frac{(W(\mu)-1)^2}{W(\mu)} \ee
\section{Derivation of Corollary \ref{cor:lebpenbound} and comparisons with previous results}
\label{sec: oldbounds}
\subsection{Derivation of Corollary \ref{cor:lebpenbound}}
We have the bound on cluster expansions from \cite{R69} as:
\be |nb_n| \leq \frac{n^{n-1}}{n!}e^{2\beta B(n-2)}C(\beta)^{-1} \ee
Where, the parameter $C(\beta)< \infty$ defines a tempered potential, which is defined in \eqref{Cbeta}, and $B$ is the stability parameter from Assumption \ref{ass: stability}. 
This corresponds to setting the parameters in the previous section to:
\begin{align} a &= C(\beta)^{-1}e^{-4\beta B} \\
b &= e^{2\beta B}C(\beta) \\
ab &= e^{-2\beta B} \\
\mu = \frac{eab}{1+ab} &= \frac{e}{1+e^{2\beta B}} 
\end{align}
This gives:
\be |c_n| \leq \frac{\beta^{-1}}{n}\left(C(\beta)e^{4\beta B} \frac{W\left(\frac{e}{1+e^{2\beta B}}\right)}{(W\left(\frac{e}{1+e^{2\beta B}}\right)-1)^2}\right)^{n-1} \ee
with the radius of convergence satisfying:
\be \mathcal{R}_{\text{Vir}} \geq C(\beta)^{-1}e^{-4\beta B}\frac{(W\left(\frac{e}{1+e^{2\beta B}}\right)-1)^2}{W\left(\frac{e}{1+e^{2\beta B}}\right)} \label{firstbound} \ee
\subsection{Morais-Procacci Bound}
This is an alternative derivation of the Morais-Procacci bounds using the method outlined in Section \ref{sec: general}.

In the recent paper \cite{MP13}, Morais and Procacci obtain a bound for the virial coefficients via the Canonical Ensemble, using polymer expansion methods and the bounds for cluster coefficients, in the form:
\be |nb_n| \leq \frac{n^{n-1}}{n!}e^{2\beta B(n-2)}C(\beta)^{-1} \ee

The values $a=C(\beta)^{-1}e^{-4\beta B}$; $b=e^{2 \beta B}C{\beta}$; and thus $ab=e^{-2\beta B}$ are substituted into \eqref{reqn}, so that we get:
\be r(s)=s(e^{-s}(1+e^{- 2\beta B})C(\beta)^{-1}e^{-2 \beta B}-C(\beta)^{-1}e^{-4\beta B}) \label{proceqn} \ee
If we let $u=e^{2\beta B}$, then our equation becomes:
\be r(s)=C(\beta)^{-1}\frac{1}{u}s(e^{-s}(1+\frac{1}{u})-\frac{1}{u}) \label{uform} \ee
We reiterate the remark that $s \in (0, \ln (1+u))$ so that we change variables (monotonically) to $s = \ln (1+u(1-e^{-\alpha}))$, so that:

\begin{align}
\tilde{r}(\alpha)&=\frac{1}{C(\beta)u}\left(\ln(1+u(1-e^{-\alpha}))\left(\frac{u+1}{u(1+u(1-e^{-\alpha}))}-\frac{1}{u}\right)\right) \label{tildeeqn} \\
&=  \frac{1}{C(\beta)u} \frac{\ln(1+u(1-e^{-\alpha}))}{e^{\alpha}(1+u(1-e^{-\alpha}))}
\end{align}
This then gives us the bound for $|\rho|$ as:
\be |\rho| \geq C(\beta)^{-1}\max_{\alpha \in (0, \infty)} \frac{\ln(1+u(1-e^{-\alpha}))}{ue^{\alpha}(1+u(1-e^{-\alpha}))} \label{alphabeta} \ee
If we then follow the same argument before, we get the bound for the virial coefficient as:
\be |c_n| \leq \frac{\beta^{-1}}{n}C(\beta)^{n-1} (\mathcal{F}(u))^{-(n-1)} \ee
where $\mathcal{F}(u)=\max_{a \in (0, \infty)} \frac{\ln(1+u(1-e^{-\alpha}))}{ue^{\alpha}(1+u(1-e^{-\alpha}))}$. 

The main improvement in the paper comes from better techniques of approximating virial coefficients. Using techniques involving Canonical Ensemble calculations for free energy and its relationship as the Legendre transform of pressure, they obtain slightly better bounds on the virial coefficients, although still under the proviso that $|\rho| \leq \rho^*$, where $\rho^*$ is the radius of convergence in \eqref{firstbound}. The improved estimates for the coefficients seem to imply that there may be a way of extending the radius of convergence, at least for temperatures for which the bound on the coefficient is an improvement.
They obtain bounds on the coefficient of $\rho^{k+1}$ in the free energy as:
\be \left( \frac{1}{k+1}+(e^{\alpha^*_{\beta}}-1)e^{\alpha^*_{\beta}k}\right)e^{2\beta B(k-1)}\frac{(k+1)^k}{k!}C(\beta)^k \ee
where $\alpha^*_{\beta}$ is the optimal $\alpha$ from \eqref{alphabeta}. This leads to the (better) asymptotic bound for the virial coefficients as:
\be K\left(\frac{e^{2\beta B}C(\beta)}{0.24026}\right)^k \ee where $K$ is a constant.

\subsection{Lebowitz-Penrose}
We again use the same bounds on the cluster expansion and follow the method outlined in \cite{LPen64} and starting with \eqref{uform}, we arrange our expression $r(s)$ into the form:
\be r(s)=\frac{C(\beta)^{-1}}{1+u}\frac{s}{u^2}\left((1+u)^2e^{-s}-(1+u)\right) \label{lebpenstart}\ee
We use the identity: $\frac{1+u}{u^2}=\frac{(1+u)^2}{u^2}-\frac{1+u}{u} $, to rewrite it as:
\be r(s)=\frac{C(\beta)^{-1}}{1+u}\left(s\frac{1+u}{u}-s^2\frac{(1+u)^2}{u^2}\left(\frac{1-e^{-s}}{s}\right)\right) \ee
We make the change of variables $v= s\frac{1+u}{u}$ and define $g(w):=\frac{1-e^{-w}}{w}$ and obtain:
\be \hat{r}(s)=\frac{C(\beta)^{-1}}{1+u}(v-v^2g\left(\frac{uv}{1+u}\right)) \ee
We note that $g'(w)=\frac{we^{-w}-(1-e^{-w})}{w^2}=\frac{(w+1)e^{-w}-1}{w^2}$

If we use the inequality $e^w \geq 1+w$, then $1 \geq (1+w)e^{-w}$ and so $0 \geq (1+w)e^{-w}-1$ and hence $g$ is decreasing. We note that $u \in (1,\infty)$ and so $\frac{u}{u+1} \in (\frac{1}{2},1)$. Since $g$ is decreasing, we get maximum value at $g(\frac{v}{2})$, which would give a minimum value for the expression in brackets, which gives a bound uniform in $u$. This is an extra approximation in the derivation of the virial coefficient bounds, which is not made in the general derivation. This is equivalent to realising that the $g$-function is dominated by its contribution at $u=1$, that is, when $\beta=0$, so at high temperatures.

We then seek to maximise: $f(v)=v-v^2g(\frac{1}{2}v)$.
Upon differentiation we get:
\begin{align} f'(v) &= 1-2vg(\frac{1}{2}v)-\frac{1}{2}v^2g'(\frac{1}{2}v) \notag \\ &=1-2v\left(\frac{1-e^{-\frac{1}{2}v}}{\frac{1}{2}v}\right) -\frac{1}{2}v^2\left(\frac{(\frac{1}{2}v+1)e^{-\frac{1}{2}v}-1}{\frac{1}{4}v^2}\right) \notag \\ &= 1-4(1-e^{-\frac{1}{2}v})-2(\frac{1}{2}v+1)e^{-\frac{1}{2}v}+2 \notag \\ &=-1+2e^{-\frac{1}{2}v}-ve^{-\frac{1}{2}v} \end{align}
When we set this to zero we get:
\be 1=(2-v)e^{-\frac{1}{2}v} \ee
changing parameters to $\delta=1-\frac{1}{2}v$, we have:
\be \frac{e}{2}=\delta e^{\delta} \ee
and so using the Lambert W-function again we get:
\be \delta=W\left(\frac{e}{2}\right) \ee
In our original expression:
\begin{align} \tilde{f}(\delta)&=2-2\delta-(2-2\delta)^2g(1-\delta) \notag \\ &=2(1 - \delta)-4(1-\delta)^2\left(\frac{1-e^{\delta-1}}{1-\delta}\right) \notag \\ &= (1-\delta)(2-4(1-e^{\delta}e^{-1})) \notag \\ &= 2(-1+\delta+\frac{1}{\delta}-1) \notag \\ &=2\frac{(W(\frac{e}{2})-1)^2}{W(\frac{e}{2})} \end{align}
This therefore gives us the bound 
\be |\rho| \geq \frac{2C(\beta)^{-1}}{1+u}\frac{W(\frac{e}{2})-1)^2}{W(\frac{e}{2})} \ee
Leading to the coefficient bound of:
\be |c_n| \leq \frac{\beta^{-1}}{n}C(\beta)^{n-1}(1+e^{2\beta B})^{n-1}\left(\frac{W(\frac{e}{2})}{2(W(\frac{e}{2})-1)^2}\right)^{n-1} \ee
and the lower bound on the radius of convergence as:
\be C(\beta)^{-1}\frac{1}{1+e^{2\beta B}}2\frac{(W(\frac{e}{2})-1)^2}{W(\frac{e}{2})} \label{LPenradius} \ee
\subsection{Comparison of the Bounds}
\begin{figure}[H]
\includegraphics[scale=1]{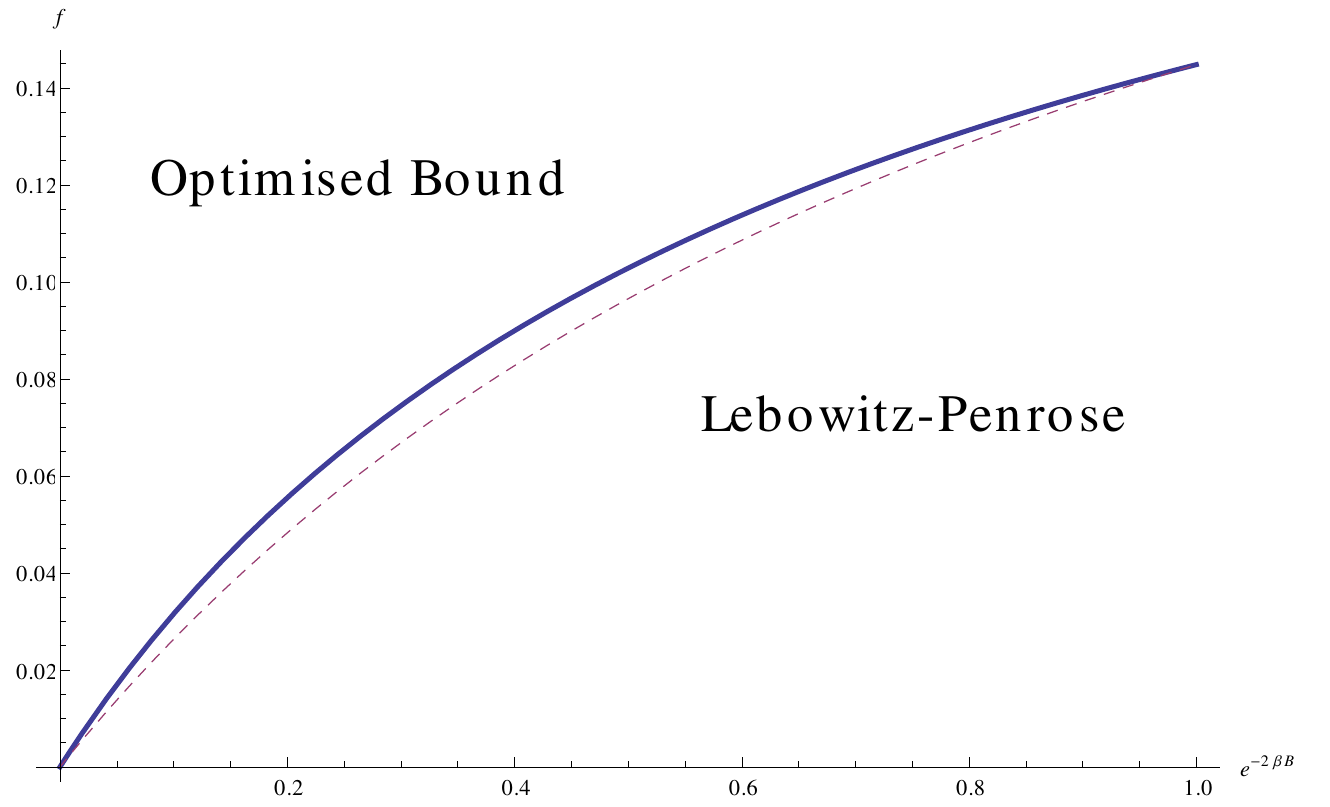}
\caption{Comparison of the Optimised Bound ($r_1$) \eqref{r1} with the Lebowitz-Penrose Bound ($r_2$) \eqref{r2}}
\label{fig:optlebpen}
\end{figure}

\begin{figure}[H]
\includegraphics[scale=1]{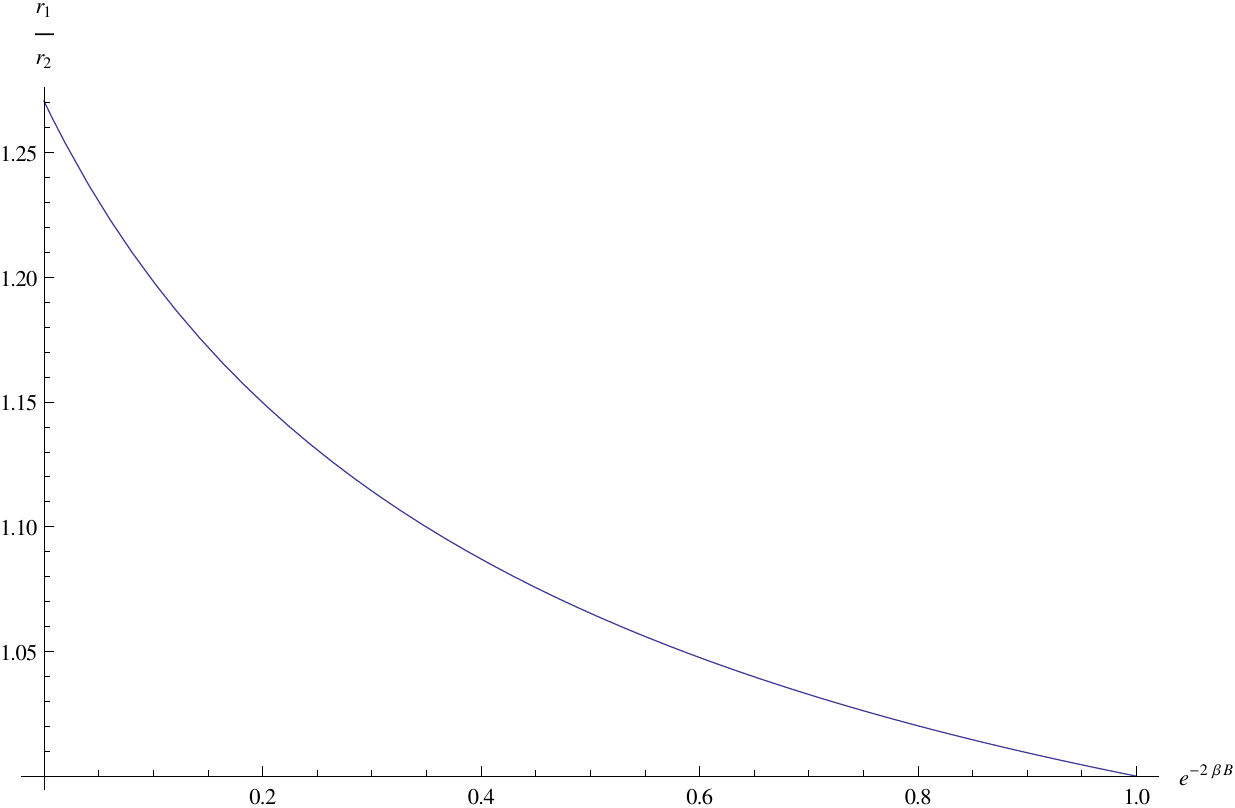}
\caption{Quotient of the Optimised Bound ($r_1$) \eqref{r1} and Lebowitz Penrose Bound ($r_2$) \eqref{r2}}
\label{fig:lebpenquot}
\end{figure}

If we define:
\be r_1:= e^{-4\beta B}\frac{(W\left(\frac{e}{1+e^{2\beta B}}\right)-1)^2}{W\left(\frac{e}{1+e^{2\beta B}}\right)} \label{r1}\ee
\be r_2:= \frac{1}{1+e^{2\beta B}}2\frac{(W(\frac{e}{2})-1)^2}{W(\frac{e}{2})} \label{r2} \ee
In order to concentrate on the factor over which the two bounds \eqref{LPenradius} and \eqref{firstbound} differ, we need only concentrate on $r_1$ and $r_2$ above. In Figure \ref{fig:optlebpen}, we see that the optimised bound shows a slight improvement over the Lebowitz-Penrose bound. Furthermore, considering the quotient $\frac{r_1}{r_2}$ in Figure \ref{fig:lebpenquot}, we see that the optimised bound is $1.25$ times better for the low temperature limit, whereas at high temperatures the two bounds are approximately the same. This is explained by emphasising that the approximation on the $g$-function is its exact value at zero $\beta$ or high temperature. 
\section{The Alternative Bound: Corollary \ref{cor:mybounds}}
There are alternative cluster coefficient bounds relating to Tree-Graph-Identities found in \cite{B86}, \cite{P07} and \cite{PU09}.
The bounds are given by:
\be |nb_n| \leq \frac{n^{n-1}}{n!}R(\beta)^{n-1}e^{n\beta B} \label{altbounds} \ee
where \be R(\beta)= \left(|\mathbb{B}|r^d+\beta \int\limits_{|y|>r}\, |\Phi(0,y)| \mathrm{d}^dy \right) \label{Reqn} \ee
where $|\mathbb{B}|$ denotes the surface area of the unit sphere in $d$-dimensions. In this case our parameters are: $a=R(\beta)^{-1}$; $b=R(\beta)e^{\beta B}$; $ab=e^{\beta B}$; and $\mu=\frac{e^{\beta B +1}}{1+e^{\beta B}}$.

The general bound for $|\rho|$ is (from \eqref{finalreqn}):
\be |\rho| \leq R(\beta)^{-1}\frac{(W(\frac{e^{\beta B +1}}{1+e^{\beta B}})-1)^2}{W(\frac{e^{\beta B +1}}{1+e^{\beta B}})} \label{altboundsrho} \ee
Which gives the bound on the coefficients as:
\be |c_n| \leq \frac{\beta^{-1}}{n}R(\beta)^{n-1}\left(\frac{W\left(\frac{e^{\beta B +1}}{1+e^{\beta B}}\right)}{\left(W\left(\frac{e^{\beta B+1}}{1+e^{\beta B}}\right)-1\right)^2}\right)^{n-1} \ee
and the lower bound on the radius of convergence is:
\be R(\beta)^{-1}\frac{(W\left(\frac{e^{\beta B+1}}{1+e^{\beta B}}\right)-1)^2}{W\left(\frac{e^{\beta B+1}}{1+e^{\beta B}}\right)} \ee
\section{Comparison of the Separate Bound: Proposition \ref{prop: comparison}}
\label{sec: comparison}
\begin{figure}[H]
\includegraphics[scale=1]{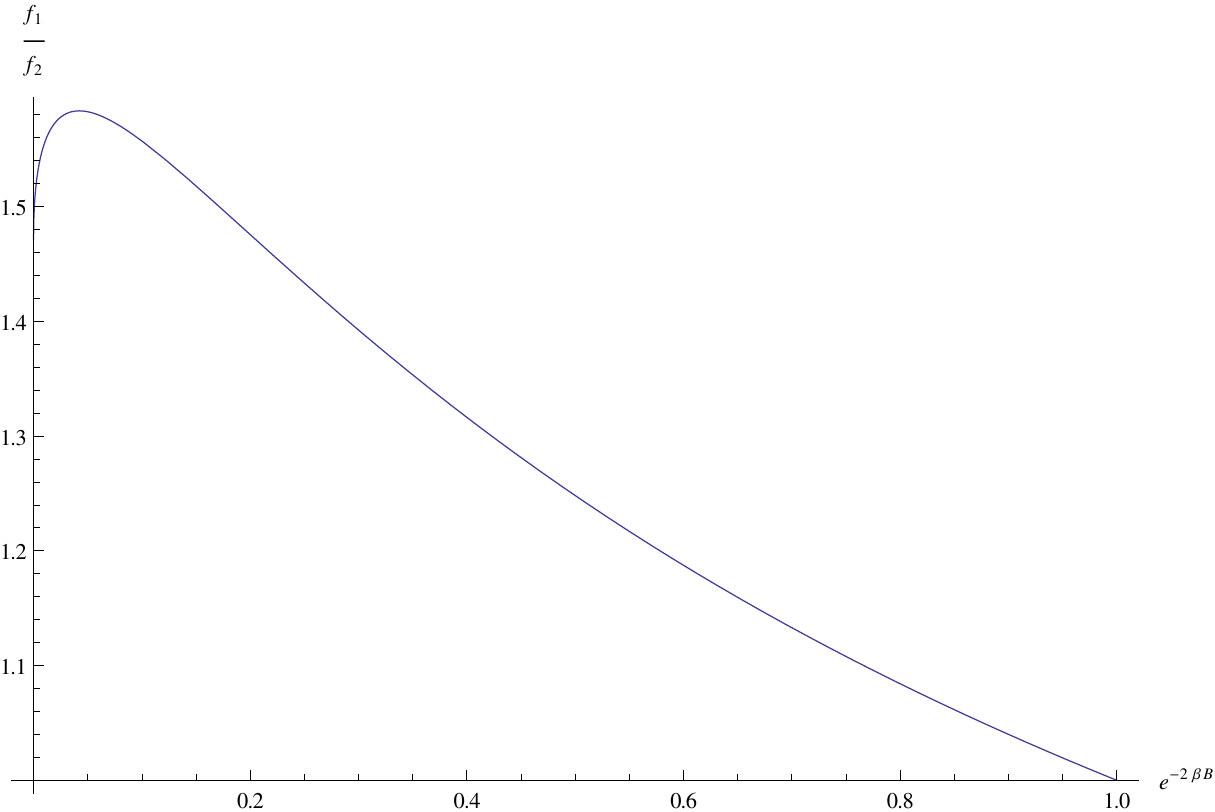}
\caption{Quotient $\frac{f_1}{f_2}$ \eqref{compare2} of the Optimised Penrose-Lebowitz Bound ($f_1$) \eqref{f1} over the New Bound ($f_2$) \eqref{f2} }
\label{fig:quotient}
\end{figure}

This approach would be better for potentials where $\zeta R(\beta) < C(\beta)$, for some $\zeta$ representing the quotient of the coefficients $\frac{f_1}{f_2}$, where:
\begin{align} f_1&:=e^{-4\beta B}\frac{\left(W\left(\frac{e}{1+e^{2\beta B}}\right)-1\right)^2}{W\left(\frac{e}{1+e^{2\beta B}}\right)} \label{f1} \\
f_2&:=\frac{\left(W\left(\frac{e^{\beta B+1}}{1+e^{\beta B}}\right)-1\right)^2}{W\left(\frac{e^{\beta B+1}}{1+e^{\beta B}}\right)} \label{f2}\end{align}
$f_2$ is actually smaller than $f_1$, as detailed in Figure \ref{fig:quotient}. This would only be an improvement if the $R(\beta)$ factor compensates. In the case of a hard-core contribution, both $R(\beta)$ and $C(\beta)$ are the same, and so are the $f_i$ factors. The use of the integrand $\beta |U(y)|$ rather than $|e^{-\beta U(y)}-1|$ is better for potentials with attractive parts. For $\beta=0$, the coefficients $f_1$ and $f_2$ are precisely the same.

The comparison between the optimised bound in \eqref{f1} and the bound achieved from these different cluster coefficient bounds based on tree graph identities ($f_2$) is shown in Figure \ref{fig:quotient}. This gives us that as soon as $1.6 R(\beta) < C(\beta)$ this other version of bounds is better. This is due to the fact that 
\be \frac{C(\beta)}{R(\beta)}> 1.6> \frac{f_1}{f_2} \label{compare2} \ee 
and so 
\be f_2 R(\beta)^{-1} > f_1 C(\beta)^{-1} \ee

We also note that $R(\beta)$ should be a better bound, for large $\beta$ or small temperature, since it is linear in $\beta$, whereas $C(\beta)$ is exponential in $\beta$ and that this $\zeta(\beta)$ could be understood better to explore the comparison of these two bounds. It also only depends on the potential $\Phi$ through the stability parameter $B$.

\appendix
\section{Lagrange Inversion}
To find the inverse of the function $se^{-s}$ as a power series, we can use the one-dimensional Lagrange inversion form from \cite{MSV06}. We want to find a power series of $s$ in terms of $y=se^{-s}$, which amounts to finding the inverse. We write the equation in the convenient form:
\be s=ye^s \ee
For a formal power series $s(y)$, implicitly defined by $s=y\phi(s)$, the Lagrange Inversion Formula can be stated as:
\be [y^n]s = \frac{1}{n} [s^{n-1}]\phi(s)^n \ee
 
where the notation $[y^n]f(y)$ is used to denote the coefficient of $y^n$ in the expression $f(y)$.

The function $\phi$ on which we wish to perform Lagrange Inversion is $\phi(s)=e^s$ and the formula for the $n$th coefficient in the desired power series is:
\begin{align} [y^n]s &=\frac{1}{n}[s^{n-1}]e^{ns} \notag \\
&=\frac{1}{n} \frac{n^{n-1}}{(n-1)!} = \frac{n^{n-1}}{n!} 
\end{align}
The power series we obtain is $s = \sum\limits_{n \geq 1} \frac{n^{n-1}}{n!}y^n$ as required.

\medskip\noindent
{\bf Acknowledgements.}
I would like to thank my supervisor Dr. Daniel Ueltschi and Dr. Sabine Jansen for helpful comments during the preparation of this paper. I would also like to thank the referees for helpful comments in the final editing of the paper.

\end{document}